\documentstyle[12pt]{article}
\begin{document}
% \draft
%\renewcommand{\baselinestretch}{1.5}
\title{Spontaneous Symmetry Breaking as the Mechanism of Quantum
Measurement}
\author{Michael Grady\\
Department of Physics\\ SUNY Fredonia, Fredonia, NY 14063 USA}
\date{}
\maketitle
\thispagestyle{empty}
\begin{abstract}
It is proposed that an event that constitutes a
quantum measurement corresponds to the spontaneous breaking of
a symmetry in the measuring device over time.
\end{abstract}
%\pacs{xxx}
\newpage
The mechanism by which a quantum measurement is performed
has been the subject of much controversy. The major problem is
the lack of a precise mathematical description of the measurement
process. If one insists on measuring devices themselves obeying
quantum mechanics and evolving according to, say, the Schr\"odinger
equation, then it is difficult to see how a measurement actually
takes place, since a measuring device coupled to a system in a
superposition of two states will itself end up in a superposition.
Thus it is necessary to add an ad hoc measurement hypothesis to
the ``rules of quantum mechanics'' whereby a measuring device is
endowed with the special property of being able to break the
superposition between states it is capable of measuring, and immediately
after the measurement is only allowed (along with the system)
to exist in an eigenstate
of the measured quantity. Such an hypothesis works in practice to
explain the results of experiments, but the description seems incomplete
in that there is no mathematical description of the evolution of the
measuring device and the system being measured
from its state before to its state after the
measurement,
leaving open puzzles such as the exact process and space-time pattern
of wave function collapse\cite{AA}.
In addition, there seems no completely satisfactory relativistic
generalization of measurement theory.

This basic problem in measurement theory is succinctly exemplified
by the ``Schr\"{o}dinger's Cat'' paradox, where the absurdity of a
measuring device
remaining in a superposition is taken to the extreme by having a cat
in a superposition of alive and dead states\cite{SC}. I think that most
people
would agree that it is absurd to think that the cat's fate has not
already been decided before one looks into the box.

Thus one is led to look for some reason by which one
can prohibit measuring devices from being in superpositions of the
quantities they are designed to measure. It is reasonable to assume
that at least the individual components (i.e. atoms) which make up
the measuring device still obey ordinary quantum mechanics. One idea
is that since measuring devices tend to be large and contain large
numbers of degrees of freedom, they are subject to random
thermal fluctuations which will randomize those phases that would
be necessary to establish and sustain such a superposition. For
instance,
it has been suggested that it is impossible to isolate a measuring
device from its environment allowing for such random influences to enter
from the outside\cite{Zurek}. Although this provides a possible
explanation,
it is somewhat unsatisfying in that by arguing that complexity is a
basic
requirement of a measuring device it would seem that any description of
the measuring process would also be necessarily rather complex. It also
does not provide a sharp dividing line between ``ordinary quantum
systems''
which can exist in superpositions and ``measuring devices'' which
cannot.	 Another possibility suggested recently involves decorrelation
through coupling to higher modes in a string theory\cite{Nano}.

In the following it is suggested that what distinguishes measuring
devices is that
they contain spontaneously broken symmetries, the
generators
of which relate the eigenstates of the measured quantity. Not only this,
but the process of measurement corresponds to the actual breaking of the
symmetry as time evolves. The similarity between the measurement process
and spontaneous symmetry breaking has been previously
pointed out by Ne'eman\cite{Neeman}. Anderson has also suggested a
connection
in that measuring devices generally consist of rigid materials which
must contain broken symmetries\cite{Anderson}. Here it is suggested that
they are one and the same and the ideas are expanded upon and developed.

What is needed is not just a system with a broken symmetry but one with
an {\em adjustable parameter} which can take it from the symmetric to
the
broken phase. The measuring device is first coupled to the system being
measured while in the symmetric phase. In this phase the measuring
device
attains whatever superposition the system being measured is in. Then
the symmetry breaking parameter is adjusted until the symmetry breaks,
throwing the entire combined system into one or another broken
eigenstate.
The superposition is broken due to the non-ergodicity of the broken
ground states of the measuring device. In quantum
mechanical language this
translates into the existence of a
superselection operator which selects
for and separates the different broken
ground states. By having a time
dependent symmetry breaking parameter, the existence or non-existence
of such a superselection operator is also made time dependent. Sherry
and
Sudarshan have previously utilized superselection operators in an effort
to explain the lack of superposition in measuring devices\cite{SS}.
If enough
superselection operators are present, the system is found to behave
essentially
classically. However, in order to perform a measurement, the
superselection
operators must be turned off while the measuring device is coupled to
the quantum system being measured. Here it is argued that breaking
and unbreaking of a symmetry spontaneously provides such a mechanism for
effectively turning on and off superselection operators.

As an example, imagine a device designed to measure the spin of a
particle.
The measuring device is represented by a scalar field theory based on a
real scalar field, $\phi$, with Lagrangian density
\begin{equation}
\frac{1}{2} \partial _\mu \phi \partial ^\mu \phi
-\frac{1}{2} \mu ^2 \phi ^2 - \frac{1}{4} \lambda \phi ^4
+ \epsilon \phi \overline{\psi} \sigma _i \psi
\end{equation}
where $\sigma _i$ is a Pauli spin matrix in the non-relativistic case
or $\sigma_i = \sigma _{\mu \nu}$ for a relativistic treatment.
For this problem it will be assumed that there is known to be one
fermion
with creation operator $\psi ^\dagger$ located somewhere within the
interaction region, which has
been prepared in a state which is a superposition
of eigenstates of $\sigma _i$, the spin component to be measured.
Thus an explicit symmetry breaking term is
presented to the scalar field, the sign of which is determined by
the spin of the particle being measured. One begins at time $t = -T$
with $\epsilon=0$, $\mu ^2 > 0$. Then the systems are coupled by setting
$\epsilon > 0$. Finally $\mu ^2$ is reduced until
it becomes negative (at say $t=0$) at which point the symmetry breaks.
Before $\mu ^2$ becomes negative, configurations with
$\overline{\phi} > 0$ and $\overline{\phi} < 0$ (where $\overline{\phi}$
represents the spatial average of $\phi$) both exist, which are somewhat
correlated with the spin-up and spin-down states of the fermion.
The fact that both exist means that a superposition is present.
As $\mu^2 \rightarrow 0$ the system becomes more and more sensitive
to the explicit symmetry breaking. Finally, for $\mu ^2 < 0$
configurations are seperated into two distinct ensembles representing
the two broken ground states, with no tunneling between them.
At this point the superposition is broken and the system can be said
to be definitely in one state or the other, i.e. the system is now in a
mixed state. Why? Because superpositions of spontaneously broken
vacuum states are not allowed states in field theory. Such states
violate
the cluster property, for instance.
To see this consider a hypothetical symmetric superposition of
two broken vacua, one with
$<\! \phi \! >\, = v$ and the other with $<\! \phi \! >\, = -v$.
This superposition will have
$<\! \phi\! >\, = 0$, but $\lim_{x \rightarrow \infty}
<\! \phi (0) \phi (x) \! >\, = v^2$.
{}From the path integral point of
view one is supposed to no longer sum over all configurations but
only over one subset corresponding to a given broken vacuum,
and excitations
around it.

Technically, these ensembles are only completely ergodically isolated
if the measuring device contains an infinite number of degrees of
freedom.
Real measuring devices contain a large but finite number of degrees of
freedom. However, if the tunneling time is very large, say many times
the age of the universe, it seems reasonable to assume that the
behavior will be the same as that of an infinite system. Very much
the same issue arises in early universe phase transitions. Consider
the electro-weak phase transition, for example. Here it is crucial
that a specific symmetry broken vacuum arises, rather than a
superposition of different broken vacua, in order to leave the photon
massless.  One needs a single component of the Higgs
field to condense.
This must be the case even if the universe is not infinite,
and is assumed to be the case for both infinite and finite universes.

The identification of a measuring event with spontaneous symmetry
breaking opens the door to a mathematical description of the actual
measurement process. It is the same as the dynamical breaking of a
symmetry over time. This is a well posed, but perhaps not completely
understood problem in quantum field theory. A complete understanding
of this problem  could explain the pattern of wave-function collapse
even in the relativistic case. All that is needed for this is to
use a relativistic quantum field theory to represent the measuring
device. In this picture the wave function collapse apparently
occurs along
the hypersurface on which $\mu ^2=0$, i.e. the the hypersurface
on which the symmetry breaking takes place.

As stated previously, a time dependent symmetry breaking can be
expressed
in the path integral description by letting $\mu ^2$ be a function of
time ($\mu ^2 > 0$ for $t < 0$ and $\mu ^2 < 0$ for $t > 0$). Since the
$t > 0$ portion is still infinite (or nearly infinite) the symmetry is
broken, even though it is not broken over the entire space-time. It is
clear
that the exact state immediately prior to $t = 0$ will be strongly
correlated
with the various possible broken states for $t > 0$, however all of the
sub-ensembles will look very similar for $t<<0$. This shows how
essentially
similar initial states can evolve into distinct and disjoint final
states.
The symmetry breaking chooses one of several possible histories for the
system as it dynamically chooses its ground state.
Thus the proper description of a physical situation which
includes a measurement
event contains only a subset of all possible states, those that are
included
in a single $t>0$ broken ground state, which carries with it, through
correlations across the $t=0$ hypersurface, a subset of possible
preexisting
states which nevertheless connect to nearly all possible states of the
spatial wave function at large negative times, i.e. a superposition,
consistent with the
state's preparation (initial time boundary condition).
An analogous situation which aids in visualization of this picture is
that of a ferromagnet in a spatial
temperature gradient so that $T > T_c$
(here $T$ is the temperature and $T_c$ is the critical temperature)
for $x < 0$
and $T < T_c $ for $x > 0$. The right half of the magnet will be
spontaneously magnetized and, since by itself it is an infinite
system there will be no tunneling. However the left half will be
unmagnetized, i.e. all possible magnetizations will be included in
the ensemble.

It is interesting to consider whether real measuring devices are
amenable to such a description. The bubble or cloud chamber for
instance contains a fairly obvious symmetry breaking as an essential
ingredient of its operation. The supersaturated vapor or supercooled
liquid is a translationally invariant
metastable state lying on top of
multiple ground states consisting of condensed droplets
or bubbles, which can occur at any place in the spatial volume.
Individual ground states break translational invariance. The coupling
of the medium to the electric charge leads to a correlation between
particle position and droplet position
which influences the choice of broken
vacuum. The localization of the droplet which occurs when the
symmetry breaks carries with it the localization of the wave function
of the charged particle because the interaction causes a high
correlation
between the two positions, and since a superposition of bubbles in two
different positions is not an allowed state after symmetry breaking,
neither will be a superposition of the correlated particle positions.
Although it may be less obvious in other systems, it is conceivable
that all measuring devices contain a symmetry which is dynamically
broken upon measurement in a similar way.

Since dynamical spontaneous symmetry breaking apparently allows a
physical system to evolve from a pure state to a mixed state,
it may also play a role in the information loss paradox present
in the formation of and subsequent evaporation of
black holes\cite{Hawking}.

Finally, such a picture leads to certain speculations as to the nature
of time in quantum mechanics and quantum field theory. The spontaneous
breaking of symmetries explains why a system has several possible
distinct histories (corresponding to a mixed state) rather than always
a superposition of all possible histories, however it does not
explain why one particular history is chosen rather than another in a
given case. In other words there is still a ``roll of the dice''.
This could be inherent to the quantum concept of reality, however if
the usual picture of a superposition is modified somewhat an element
of determinism can be introduced. One can picture a superposition
as a system which is very rapidly switching between all possible
states in the ensemble (for a field theory this would be all
possible classical field configurations
consistent with boundary conditions). Thus ordinary time would in some
sense be connected to a ``Monte Carlo'' time in such a way that a
very small interval of ordinary time involved many steps in Monte Carlo
time over which
the field configuration was randomly changed,
including past values of the field. In this case the fields
really would have a particular value at any one instant and the
direction of symmetry breaking would depend on the exact field
configuration
present at the exact moment of symmetry breaking.
However it is far from clear whether such a radical picture of time
evolution can be made consistent with fundamental properties
such as relativistic covariance and causality.

It has been suggested that spontaneous symmetry breaking events
constitute quantum measurements. Because consistency of field theory
prohibits superposition of different spontaneous broken vacuum states,
the dynamical breaking of a symmetry over time results in the
transformation
of a pure state into a mixed state. This will break the superposition of
any quantum state coupled strongly enough to the broken symmetry
generator.
Thus measuring devices are a special class of quantum systems which
contain an adjustable symmetry breaking parameter which can take the
system into and out of the spontaneously broken state.

\begin{center}
Acknowledgement
\end{center}
It is a pleasure to acknowledge the hospitality of
the Theoretical Physics Group at the Tata Institute of
Fundamental Research, where a portion of this work was completed.
This work was partially supported be a grant from the Indo-US
Subcommission on Education and Culture.

\end{document}